\documentclass[journal=aamick,manuscript=article]{achemso}

\pdfoutput=1 
\usepackage[version=3]{mhchem}

\usepackage{chemformula} 
\usepackage[T1]{fontenc} 
\usepackage{amsmath,amsfonts,amssymb}
\usepackage{graphicx} 
\usepackage{subcaption} 
\usepackage[separate-uncertainty = true, multi-part-units = single, list-units = single, range-units = single]{siunitx} 
\DeclareSIUnit\Molar{\textsc{m}}
\DeclareSIUnit\rpm{rpm}
\DeclareSIUnit\ppm{ppm}

\DeclareSIUnit\kbt{k_BT}

\usepackage[noabbrev,nameinlink]{cleveref}
\usepackage{multirow}
\usepackage{makecell}
\usepackage{colortbl}
\usepackage{upgreek}
\usepackage{placeins}

\title{Assembly pathways of anisotropic lipid membrane-deforming colloids}

\author{Ali Azadbakht}
\affiliation[Leiden University]{Soft Matter Physics, Huygens-Kamerlingh Onnes Laboratory, Leiden University, PO Box 9504, 2300 RA Leiden, the Netherlands}

\author{Thomas Weikl}
\affiliation[MPI]{Max Planck Institute of Colloids and Interfaces, Am M\"uhlenberg 1, 14476 Potsdam, Germany}
\email{thomas.weikl@mpikg.mpg.de}

\author{Daniela J. Kraft}
\affiliation[Leiden University]{Soft Matter Physics, Huygens-Kamerlingh Onnes Laboratory, Leiden University, PO Box 9504, 2300 RA Leiden, the Netherlands}
\email{kraft@physics.leidenuniv.nl}

\begin{document}
\begin{abstract}
Membrane-deformation mediated interactions play an important role in the spatial organization of proteins on the cell membrane. Although interactions between isotropic membrane deformations have been extensively investigated, the role of anisotropic deformations remains largely unexplored despite their prevalence in biological systems.
Here, we experimentally investigate the assembly of anisotropic colloidal objects that deform a lipid membrane while being confined underneath it, without direct attachment. 
Combining experiments and numerical calculations, we analyze how a wide range of shapes, including ellipsoids, dumbbells, cubes, scalene triangles, tetrahedra, and bent rods, interact with each other through the membrane deformations they induce.  
We find that membrane-deforming objects initially attract through regions of highest curvature and subsequently reorient into close packed arrangements with an approximately spherical circumference. 
This is achieved through the alignment of flat faces - if possible in register - and locally optimized geometric packing, with regions of high curvature imposing energy barriers that influence the assembly pathway. 
Our work reveals general principles how anisotropic membrane deformations govern the assembly pathways and final particle arrangements, providing new insights into the behavior of membrane-deforming proteins and other inclusions. 
\end{abstract}

\bigskip
\section*{Main text}
The ability of proteins to sense, generate, and respond to membrane curvature underlies many fundamental cellular processes. The diverse shapes of these proteins generate membrane deformations that have been predicted to guide their spatial organization and collective function.\cite{McMahon2005,Zimmerberg2006,Idema2019a, kozlovGenerationNanoscopicMembrane2023}
Examples include the bent rod-shape of the BAR-domain proteins, the triskelion architecture of clathrin assemblies, and the spiral filaments of ESCRT-III complexes, whose distinct geometries are intimately linked to membrane shaping, invagination, budding, and abscission.\cite{Dasgupta2014,Saleem2015,Chiaruttini2015, simunovicCurvingCellsOut2019,pfitznerESCRTIIIPolymerizationSequence2020}. 

Beyond remodeling membranes, these proteins can also interact indirectly through the membrane deformations they generate.\cite{McMahon2005,Franceschi2022}
For example, experiments with BAR domain proteins and their DNA-based synthetic models as well as coarse-grained simulations have revealed that the curvature induced by their bent-rod shape can drive directional interactions and side-by-side arrangements. At sufficiently high concentrations, these interactions promote the collective formation of membrane buds and tubular structures.~\cite{Schweitzer2015, Simunovic2013,Ramakrishnan2013,Simunovic2015b, Olinger2016,Noguchi2017,Franquelim2018MembraneScaffolds,Noguchi2017a,Bonazzi2019, Bonazzi2025}.
Directly resolving the interaction pathways and assembly dynamics of individual anisotropic proteins remains challenging because their nanometer-scale dimensions preclude \textit{in situ} observation.
In addition, proteins are subject to a multitude of biochemical and physical interactions, making it difficult to isolate the contribution arising solely from membrane deformations.
Quantitative predictions of precisely how their shape-induced membrane deformations lead to interactions and assembly are scarce, because the equations to describe this process are difficult to solve analytically. 
Yet anisotropic deformations are expected to produce directional interactions that depend on the relative orientations of the interacting objects, potentially giving rise to richer assembly pathways and more complex structures than for isotropic deformations possible. The principles governing such anisotropic deformation-mediated assembly, however, remain essentially unexplored.

Scaled-up model systems based on giant unilamellar vesicles (GUVs) and colloidal particles can overcome many of these challenges, allowing for \textit{in situ} imaging of the dynamics and quantitative measurement of forces and interactions. They have been successfully employed to investigate the wrapping of colloidal spheres and dumbbells,\cite{Spanke2020,Spanke2022, Azadbakht2023a} measure the interactions between (partially) membrane-wrapped colloidal spheres,\cite{Koltover1999, Sarfati2016, VanDerWel2016,VanderWel2017, azadbakhtRepulsionsAttractionsMembranedeforming2025}, provide experimental proof of the non-additive nature of the interaction\cite{Azadbakht2023b,azadbakhtNonadditivityManyBodyInteractions2024} and explore collective effects\cite{azadbakhtNonadditivityManyBodyInteractions2024}. The direct visualization and quantitative capabilities of these model systems make them ideally suited platforms to uncover how anisotropic membrane deformations govern interactions and assembly between anisotropic objects.

Here, we investigate how anisotropic membrane deformations generated by particle shape direct membrane-mediated interactions and self-assembly using a recently introduced attachment-free GUV–colloid model system. 
In this model system, colloidal particles confined beneath sessile, partially deflated GUVs deform the initially flat membrane, enabling direct visualization of membrane-mediated interactions.~\cite{azadbakhtNonadditivityManyBodyInteractions2024} 
With particle fabricated by synthesis and 3D microprinting, we generate a library of anisotropic colloidal particle that includes ellipsoids, bent rods, cubes, triangles, and tetrahedra, to systematically investigate how shape influences membrane-mediated interactions.
Using a combination of experiments and simulations, we identify common principles linking particle geometry to their interaction pathways and the resulting collective arrangements. 

\section*{Results and Discussions}
We utilize a recently introduced attachment-free experimental model system to study the effect of shape in lipid membrane-mediated interactions.\cite{azadbakhtNonadditivityManyBodyInteractions2024} 
In this system, anisotropic colloidal particles are confined between slightly deflated giant unilamellar vesicles and a cover slip, where they deform the membrane while remaining free to translate and rotate in the plane. Partial deflation of the vesicles by a density and osmolarity difference creates an extended flat membrane region on a polyacrylamide-coated substrate, providing a well-defined geometry for studying particle-induced deformations. 
The GUVs were composed of 1,2-dioleoyl-sn-glycero-3-phosphocholine (DOPC) containing a 2\% w/w polyethylene glycol-2000 conjugated to 1,2-dioleoyl-sn-glycero-3-phosphoethanolamine (DOPE-PEG200) to suppress nonspecific attachment (Fig.\ref{fig:fig1}). 
The simplicity of the membrane and the micrometer size of the particles enable direct visualization of their position, orientation, and interactions with confocal microscopy in the presence of thermal fluctuations. 

Membrane deformations with different shapes were generated by positioning the colloidal particles beneath the GUV using optical tweezers, see Figure \ref{fig:fig1}A.   
Fluorescent labeling of the membrane and colloids enabled real-time confocal imaging at 30Hz (Figure \ref{fig:fig1}B,C) GUVs are presented in magenta and particles in green throughout the paper. 
Three-dimensional reconstruction confirmed that the particles were confined between the substrate and the membrane (Fig. \ref{fig:fig1}B–E), where they induced pronounced anisotropic membrane deformations. 
The membrane shape follows the particles' curvature, while being free to depart from it since the particles are not attached to the membrane. 
Because the overall configuration is metastable, the particles eventually escaped from underneath the vesicle after tens of seconds to lower their free energy (Supplementary Video 1). The duration of this process allowed us to observe the interactions and self-assembly pathways between the membrane-deforming colloids.

\begin{figure}[p]
\centering
\includegraphics[width=\textwidth]{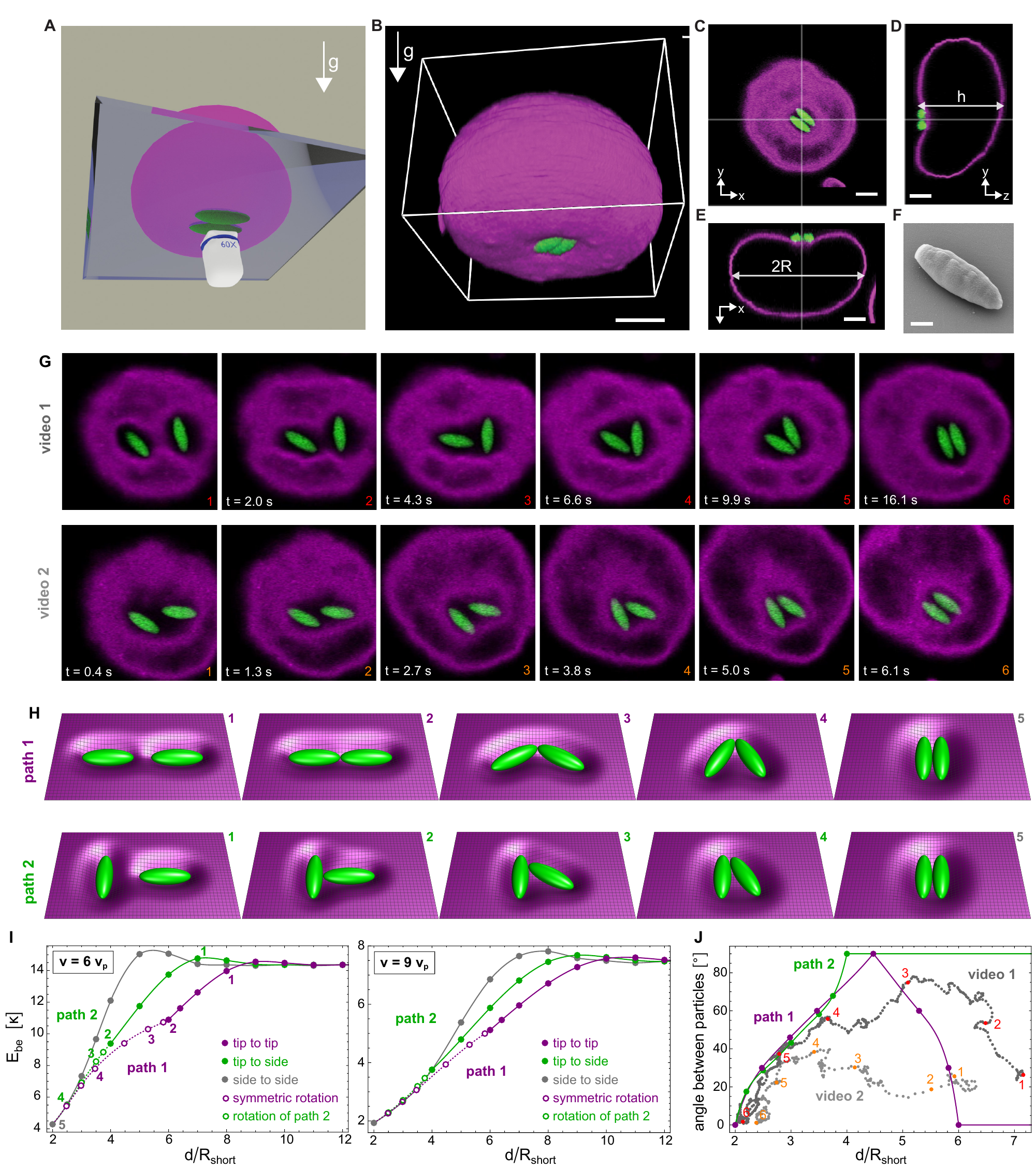}
\end{figure}
\newpage
\begin{center}
\captionsetup{type=figure}
\caption{\label{fig:fig1}\textbf{Curvature-mediated interactions of membrane-deforming ellipsoidal colloids}
(A) Experimental setup and (B) 3D confocal microscopy reconstruction of anisotropic particles (green) confined between a sessile GUV (magenta) and a cover slip; confocal microscopy observation from below. White arrow indicates the direction of gravity.
(C-E) Confocal images of (C) the plane close to the coverslip, (D) a zy-plane cross section made along the vertical line in (C), and (E) a xz-plane cross section taken along the horizontal line in (C); the GUV's height ($h$) and diameter ($2R$) are indicated; scale bars are 5~$\upmu$m;
(F) a SEM image of the 3D printed ellipsoidal colloid with 6~$\upmu$m long axis and 2~$\upmu$m short axis;  
(G) Two exemplary time series of two interacting ellipsoidal particles  (Supplementary Video 1). In both time series, the ellipsoids rotate to arrange side-by-side.
(vesicle with $h=19$~$\upmu$m and $2R=31$~$\upmu$m).
(H) Minimum-energy conformations along two exemplary pathways with decreasing particle distance $d$ for different initial orientations and final side-to-side arrangement. 
(I) Minimum membrane bending energy $E_{\rm be}$ versus distance $d$ of particle centers for different orientations of two ellipsoidal particles with 6 $\upmu$m long axis and $2 R_{\rm short}= 2 \upmu$m  short axis, for two confinement volume values.  
The minimum distances of the particles in tip-to-tip and side-to-tip orientation are 6 $\upmu$m and 4 $\upmu$m, respectively. 
(J) Particle angle versus distance as extracted from (G, dotted lines) and (H, solid lines). Red and orange numbers indicate the frames shown in (G). 
(I,J) Green and magenta numbers in indicate frames shown in (H). Solid lines are guides for the eye.
Scale bars are 2~$\upmu$m.}
\label{fig:fig1}
\end{center}

\subsection*{Particles with a smooth, convex shape: Ellipsoids}
We start exploring the influence of shape on membrane-mediated interactions using ellipsoidal colloids. The colloidal particles (long axes 6~$\upmu$m, short axes 2~$\upmu$m) were fabricated using direct laser writing (Figure \ref{fig:fig1}F, Methods for details). 
After confinement between the vesicle and substrate, the ellipsoids consistently aligned with their long axis parallel to the substrate, imposing an ellipsoidal deformation on the membrane (Fig.~\ref{fig:fig1}, Supplementary Video 1). 

To characterize their pair interaction, two ellipsoids were positioned at controlled separations and relative orientations using optical tweezers. Their center-of-mass positions and orientation were subsequently tracked over time (Fig. \ref{fig:fig1}A, Supplementary Video 1). 
Irrespective of their initial distance and orientation, the two ellipsoids reproducibly reoriented and rearranged into a side-to-side conformation with minimal center-to-center distance.  
This configuration remained stable until thermal motion brought the pair too close to the vesicle boundary, where they exited the observation region (see Supplementary Video 1, movie 2). 

To understand the robust assembly behavior and orientational changes of the ellipsoidal particles, we computed the minimum-energy membrane conformations around two confined ellipsoids using a previously established numerical approach \cite{azadbakhtNonadditivityManyBodyInteractions2024}. 
Equilibrium configurations were obtained by minimizing the non-linear bending energy subject to the geometric constraints imposed by the presence of the particles. As shown previously, the resulting interactions depends not only on the particle shape but also on the volume $v$ of the solvent pocket inside which each particle is confined \cite{azadbakhtNonadditivityManyBodyInteractions2024}. 

Fig. \ref{fig:fig1}H shows the minimum-bending-energy conformations along two representative pathways that both lead to the experimentally observed final state: a  side-to-side arrangement at the smallest possible particle distance of 2~$\upmu$m. This configuration exhibits the lowest membrane bending energy among all possible conformations.

On path 1, the particles first approach each other in tip-to-tip orientation until they come into contact at a center-to-center distance of 6 $\upmu$m. Upon further approach, the particles reorient by a symmetric rotation while remaining in contact until they reach the lowest-energy side-to-side conformation. On path 2, the particles approach each other in tip-to-side orientation until they make contact at a distance of 4 $\upmu$m. At smaller distances, one particle rotates until the particles reach the side-to-side configuration. 
Along both pathways, the rotations of the particles continuously lower the membrane bending energy $E_{\rm be}$ (Fig. \ref{fig:fig1}I).

In contrast, for particles constrained to approach in a side-to-side orientation, the membrane bending energy exhibits a pronounced maximum at an intermediate particle distance (Fig. \ref{fig:fig1}I). This energy barrier prevents further direct approach and suggests that pathways into the final side-to-side conformation require particle rotations.  

Consistent with these predictions, reorientation of the particles with decreasing particle distance is also observed in the experiments (Fig. \ref{fig:fig1}J and Supplementary Video 1). 
In Video 1, the particles initially adopt relative angles around 25$^\circ$ at distances of about 7 $\upmu$m and then reorient into a tip-to-side conformation with angles of around 75$^\circ$ at intermediate distances of about 5 $\upmu$m. At distances smaller than about 4 $\upmu$m, 
the particles are in contact in both videos and further lower their distance by rotating into side-to-side alignment with a relative angle of 0$^\circ$.

\FloatBarrier

\subsection*{Elongated particles with constant curvature and a neck: Dumbbell-shaped particles}
As a second particle shape, we investigated dumbbell colloids. These particles consist of two spherical lobes and, thus, have a constant curvature, with the exception of the highly curved neck region in the center. 
Thus, while dumbbells and ellipsoids share the overall elongated particle shape, they differ in the curvature as the ellipsoids posses a continuously varying curvature. 
Comparing these two types of particles thus allow us to investigate if an overall elongated particle shape determines the pathway of approach and final arrangement, or if the varying curvature is the dominant factor. 

\begin{figure}[htb]
\centering
\includegraphics[width=1\textwidth]{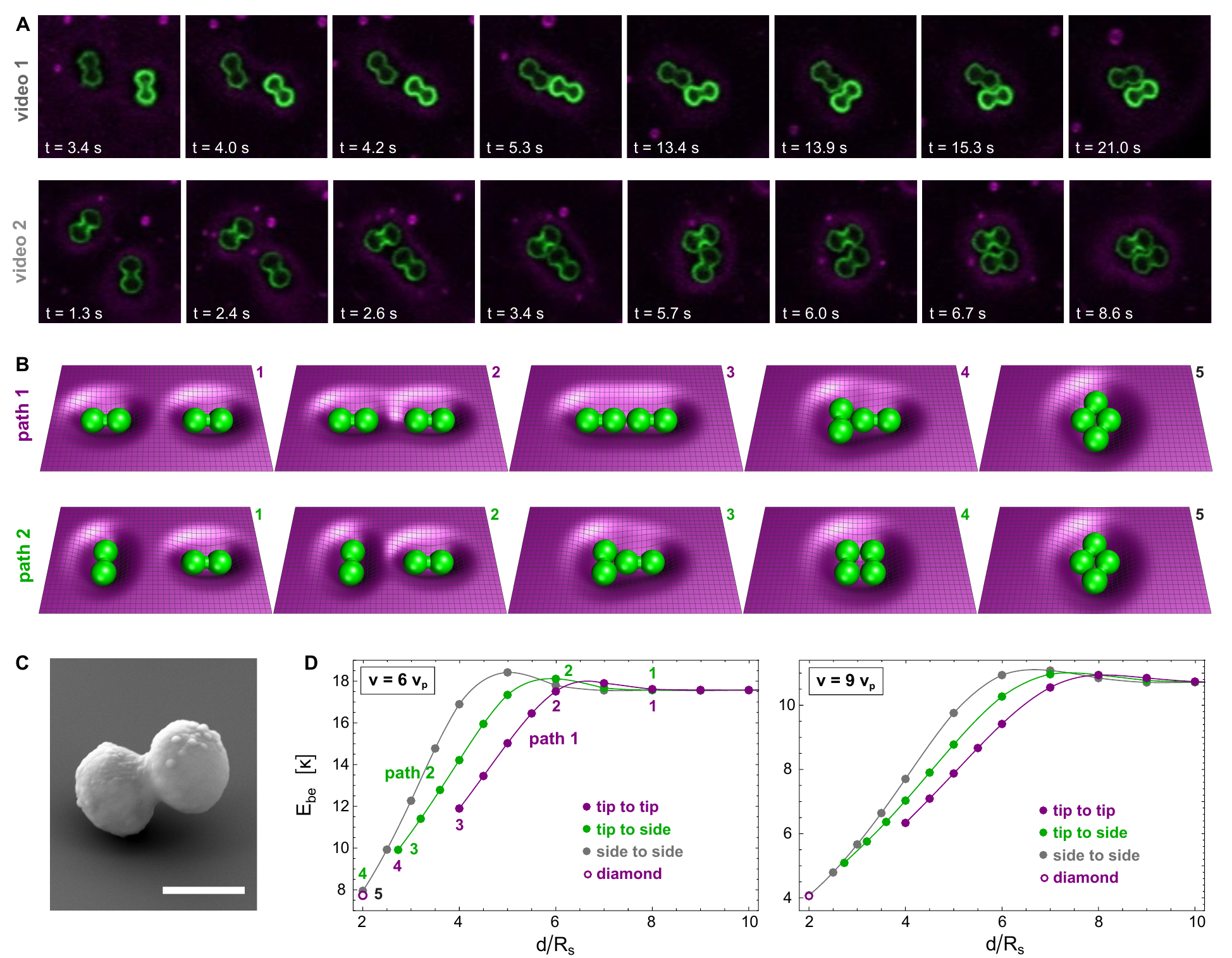}
\caption{\label{fig:dumbbells}\textbf{Curvature-mediated interaction of dumbbell-shaped colloids.} 
(A) Two representative time series of confocal microscopy images showing membrane-mediated interactions between two confined dumbbell-shaped colloids. The particles reorient during approach and ultimately arrange slightly off-set side-by-side arrangement, resembling a diamond configuration (see also Supplementary Video 2 ; vesicle with $h=19$~$\upmu$m and $2R=31$~$\upmu$m).
(B) Minimum-energy conformations along two representative assembly pathways. The particles are initially oriented tip-to-tip (path 1) and tip-to-side (path 2). Dumbbell particles were composed of two touching spheres with diameter $2 R_s = 2$ $\upmu$m. 
(C) SEM image of synthesized dumbbell particle with a length of 4~$\upmu$m and width of 2~$\upmu$m. Scale bar is 2~$\upmu$m.
(D) Minimum membrane bending energy $E_{\rm be}$ at a given distance $d$ of the particle centers and for different orientations of two dumbbell particles at the confinement volume values $v = 6 v_p$ and $9 v_p$.   
Green and magenta numbers indicate the conformations along path 1 and 2 in (B). Lines serve as guides for the eye. At the minimum distance of 2 $\upmu$m, the particles can adopt either a side-to-side conformation (4) or a ``diamond'' conformation (5). The latter is energetically slightly favourable and therefore the final conformation of both paths, in agreement with experimental observations in (A).}
\end{figure}

We employed dumbbell-shaped silica particles with dimensions of 4~$\upmu$m in the long axis and 2~$\upmu$m in the short axes, i.e. a radius of the smaller sphere $R_s=1\mu$m, see Figure~\ref{fig:dumbbells}C and Methods section for details. 
To investigate the interaction of two membrane-deforming dumbbells, we arranged them in various initial orientations and tracked their subsequent motion. 
Fig.~\ref{fig:dumbbells}A and Supplementary Video 2 show two representative time sequences of two dumbbell particles. 
When the dumbbells are initially oriented side to side (Supporting Video 2, first movie, top row of Fig.~\ref{fig:dumbbells}A, $t = 3.4$ s), they first rotate into a tip-to-tip orientation ($t = 5.3$ s). 
While remaining in contact, the particles subsequently adopt a tip-to-side orientation ($t = 13.9$ s) before reaching their final side-to-side arrangement. In this arrangement, a slight off-set allows the lobes of each particles partially fit into the waist of the other, leading to a ``diamond'' conformation. 
No further rearrangement is observed, indicating that the dumbbell particles have reached an energy minimum. 
In the second trajectory (Supplementary Video 2, second movie) the particles approach each other and get in touch in tip-to-side orientation ($t = 3.4$ s), and subsequently reorient to adopt the same diamond conformation as in the first trajectory. 

To rationalize these experimental observations, we again determine minimum-bending-energy conformations along two exemplary pathways, see Fig. \ref{fig:dumbbells}B and D.  We consider two dumbbell particles composed of spheres with diameter 2 $\upmu$m, initially oriented either tip-to-tip (path 1) and tip-to-side (path 2). 
At center-to-center distances between approximately 4 $\upmu$m and 6 $\upmu$m (equivalent to 4-6 $d/R_s$), the tip-to-tip orientation is energetically clearly favorable over tip-to-side and side-to-side orientation (see Fig. \ref{fig:dumbbells}D). 
At distances below 4 $\upmu$m, tip-to-side orientation becomes energetically favorable and remains so until a distance $(1 + \sqrt{3})$ $\upmu$m $\simeq 2.73$ $\upmu$m is reached, at which the particles are in tip-to-side contact (conformation 3 on path 2 in  Fig. \ref{fig:dumbbells}B). 
At the smallest possible distance of 2 $\upmu$m, the diamond configuration (conformation 5), in which the particles are locked in, is energeticaFlly slightly favorable compared to side-to-side alignment (conformation 4 on path 2), in agreement with the experiments (Fig. \ref{fig:dumbbells}A). 

Together, the results from the dumbbell and ellipsoid-induced membrane deformations suggests that an elongated particle shape generally promotes initial attraction along the long axis of the deformation, followed by a side-by-side arrangement. However, the neck region of the dumbbell shape allows for and favors an even more compact configuration, obtained by positioning the lobes into the other particle's waist. 

\FloatBarrier

\subsection*{Particles with flat sides and sharp corners and edges: cubes, triangles and tetrahedra}
Next, we chose to investigate particles with flat sides and sharper corners and edges. Such shapes impose flat membrane regions bounded by highly curved edges and corners that we expect to induce energetically unfavorable membrane deformations. 

\begin{figure}[ht]
\centering
\includegraphics[width=0.7\textwidth]{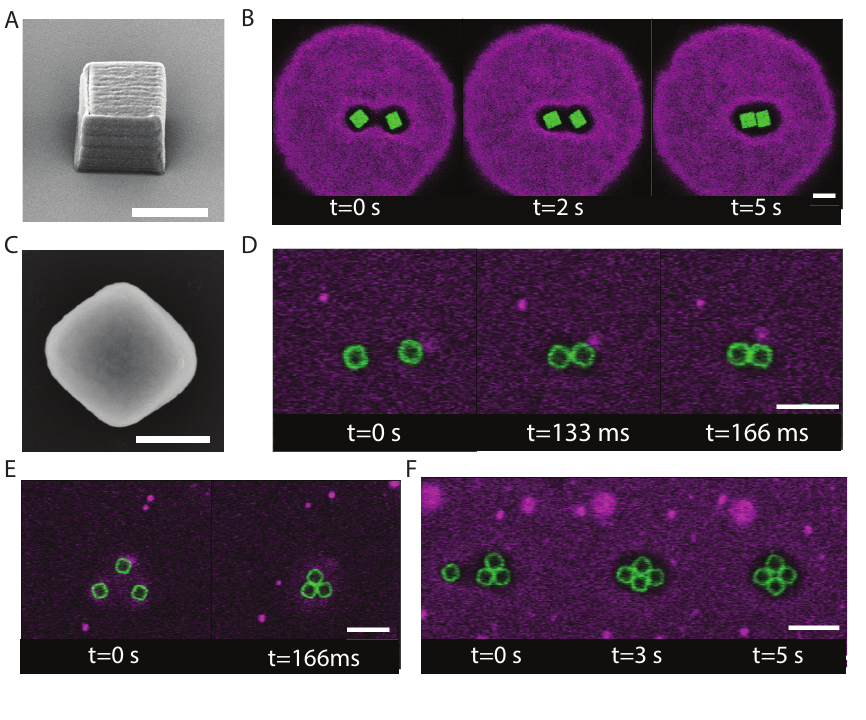}
\caption{\label{fig:cubes}\textbf{Curvature-mediated interaction of cubic colloids:} 
(A) SEM image of 3D printed cube and (B) assembly of two such cubes with sharp edges and corners through the membrane-deformation they impose.
(C) SEM image of synthesized cube with 1.6~$\upmu$m side length and (D,E,F) confocal microscopy images of the assembly of (D) two cubes (E) three, (F) four cubes through membrane deformations. 
Scale bars (A,B) 4~$\upmu$m, (C) 1~$\upmu$m, (D,E,F) 5~$\upmu$m. (B) Vesicle with $h=33$~$\upmu$m and $2R=64$~$\upmu$m (D, E,F) Vesicle with $h=60$~$\upmu$m and $2R=75$~$\upmu$m.}
\end{figure}

We started by investigating cubic particles because of their high symmetry. Two types of cubes were used: 3D-printed cubes with a side length of 4~$\upmu$m and sharp edges and corners, and synthetic silica cubes with a side length of 1.6~$\upmu$m and more rounded edges again coated with a fluorescent lipid bilayer (Fig.~\ref{fig:cubes}). 
After confinement underneath the vesicles, pairs of the larger cubes were found to rotate during approach thereby realigning their corners such that their two flat sides are in contact and perfect register (see Fig.~\ref{fig:cubes}A,B, Supplementary Video 3). The smaller cubes first touch at their corners before rotating to assume the same face-to-face configuration (Fig.~\ref{fig:cubes}C,D, and Supplementary Video 4). 
Energy minimization considerations indicate that rotations during particle approach can arise from a slight energetic preference for the edge-to-edge orientation at intermediate distances, compared to the preferred side-to-side orientation at small distances (Fig. S2).
 
While these results might suggest that the face-to-face orientation dominates the final arrangements of cubic particles, we find a more nuanced picture: rather than maximising face-to-face contact, three cubes assemble such that they touch at one of their corners and their diagonals are arranged at an angle of about 120$^\circ$ angle, see Figure~\ref{fig:cubes}D. 
This configuration is reached even when the cubes are initially aligned in a row (Supplementary Video 4). 
However, upon adding a fourth cube, the particles rearrange into a compact, square-like configuration with face-to-face contacts (Fig.~\ref{fig:cubes}E and Supplementary Video 4).

From these experiments we conclude that while face-to-face arrangements can reduce highly curved membrane regions, a compact and circular arrangement that minimizes the overall membrane deformation is favored.

\begin{figure}[ht]
\centering
\includegraphics[width=0.8\textwidth]{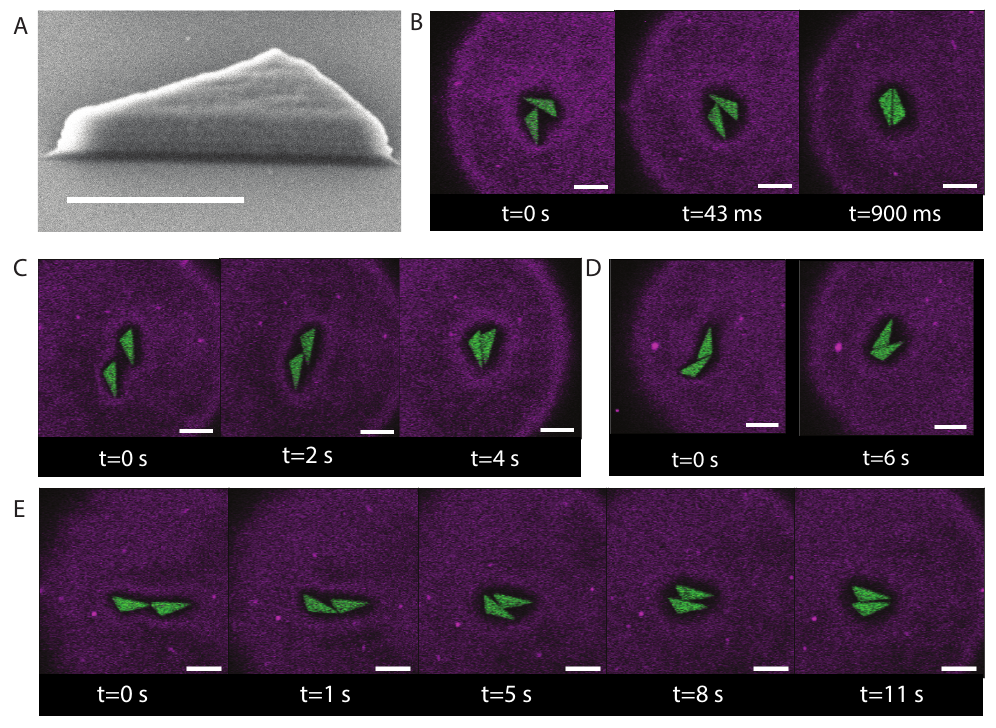}
\caption{\label{fig:triangle}\textbf{Sharp edges cause metastable states.} 
(A) SEM image of a 3D printed scalene triangular colloid; scale bar 3~$\upmu$m; 
(B, C, D, E) Four time series of confocal microscopy images showing the assembly pathways of two membrane-deforming scalene triangles starting from different initial configurations; vesicle with $h=44$~$\upmu$m and $2R=64$~$\upmu$m. See Supplementary Video 5. Scale bars 5~$\upmu$m. 
}
\end{figure}

To investigate the effect of reduced particle symmetry, we next investigated scalene triangular prisms with side lengths $a=3$, $b=5$, and $c=6.5~\upmu$m and a height of 2~$\upmu$m, as shown in Fig.\ref{fig:triangle}A,B. The sharp corners of the triangles provide a test of how local curvature influence the membrane-mediated interactions and assembly pathways.

In all observed trajectories, after first contact, triangles increased the length of their side-to-side contact by rotating and sliding relative to one another (Fig.~\ref{fig:triangle}B-E, Supplementary Video 5). For example, triangles contacting through their longest sides slid to align in perfect register (Fig.~\ref{fig:triangle}B) similar to pairs of cubes (Fig.~\ref{fig:cubes}B,D). This motion reduced the strong membrane deformations associated with the corners and drove the particles toward compact configurations. 

Unlike cubes, however, the final arrangement was not unique. We expect the configuration in which the two longest sides align to be the global energy minimum, as it minimizes both the number of exposed sharp corners and the overall cluster anisotropy. 
However, the sharp corners created substantial barriers that prevented further rearrangement by rotation, leading to kinetic trapping in metastable states. 
 Minute differences in intermediate states such as is the case in Fig.~\ref{fig:triangle}D (6s) and E (5s) then determine the final state. The final state furthermore does not provide information about the pathway, as exemplified through Fig.~\ref{fig:triangle}C and E. 
 Consequently, assembly became pathway dependent, with the final configuration determined by the initial particle orientation and location.

\FloatBarrier

\begin{figure}[ht]
\centering
\includegraphics[width=1\textwidth]{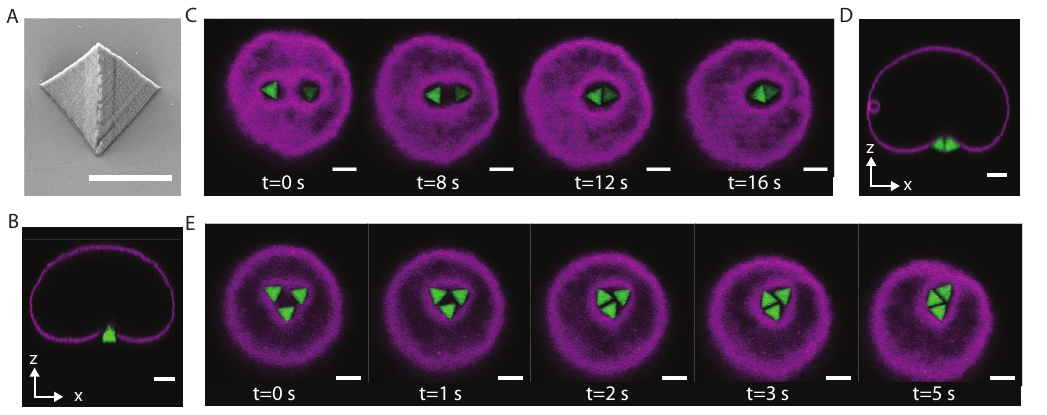}
\caption{\label{fig:tetrahedron}\textbf{Curvature-mediated interaction of tetrahedron-shape colloids:} 
(A) SEM image of a 3D printed tetrahedron. 
(B) A xz-plane cross section of a GUV deformed by a tetrahedron;
(C) Time series of confocal microscopy images of the aggregation of two membrane-deformed tetrahedra. Particles attract until two faces are in contact, moving out-of-plane to minimize vesicle deformation (Supplementary Video 6, vesicle: $h=32$~$\upmu$m and $2R=45$~$\upmu$m).
(D) $xz$-plane cross-section of (C) showing the out-of-plane arrangement of the two tetrahedra.
(E) Time series snapshots of the aggregation of three membrane-deformed tetrahedra (Supplementary Video 6, vesicle: $h=27$~$\upmu$m and $2R=34$~$\upmu$m). Scale bars 5~$\upmu$m.
}
\end{figure}

Having established some general principles for membrane-based assembly of anisotropic particles with a planar geometry, we now turn towards tetrahedra, which introduce a pronounced variation in height. Tetrahedra with an edge length of 4~$\upmu$m were fabricated by direct laser writing, see Fig.~\ref{fig:tetrahedron}A. Their high symmetry makes them the 3D counterpart of the triangular prisms and cubes studied above. 
When confined beneath a vesicle, isolated tetrahedra adopt a well-defined orientation in which one triangular face rests on the coverslip while the opposite vertex points toward the membrane (Fig.~\ref{fig:tetrahedron}B), generating a highly localized membrane deformation.

Pairs of tetrahedra attract until they touch along one of their bottom edges. Surprisingly, they subsequently rotate of the plane by lifting their bottom face away from the glass substrate until their two facets touch. See Fig.\ref{fig:tetrahedron}C, and Supplementary Video 6.  
This out-of-plane motion, confirmed by the $xz$ cross-sections in Figure \ref{fig:tetrahedron}D and 3D reconstructions shown in Fig. S3, simultaneously reduces the sharp deformation induced by their upward-pointing vertices. Given the density difference of 142 kg/m$^3$ and size, one can estimate that this lift requires about $\approx$4k$_B$T of energy which needs to be supplied from lowering the membrane bending energy.  

In contrast, three tetrahedral particles assemble into a compact linear chain with no out-of-plane lifting away from the substrate (Fig.~\ref{fig:tetrahedron}E, and Supplementary Video 6). The fact that the out-of-plane arrangement of a pair of tetrahedra does not extend to clusters of three particles suggests that the non-additive nature of membrane-deformation induced interactions comes into play, which leads to a different energetic balance with the potential energy terms.\cite{Azadbakht2023b, azadbakhtNonadditivityManyBodyInteractions2024}

\FloatBarrier

\subsection*{Shape complementarity: bent rods}
\begin{figure}[ht]
\centering
\includegraphics[width=1\textwidth]{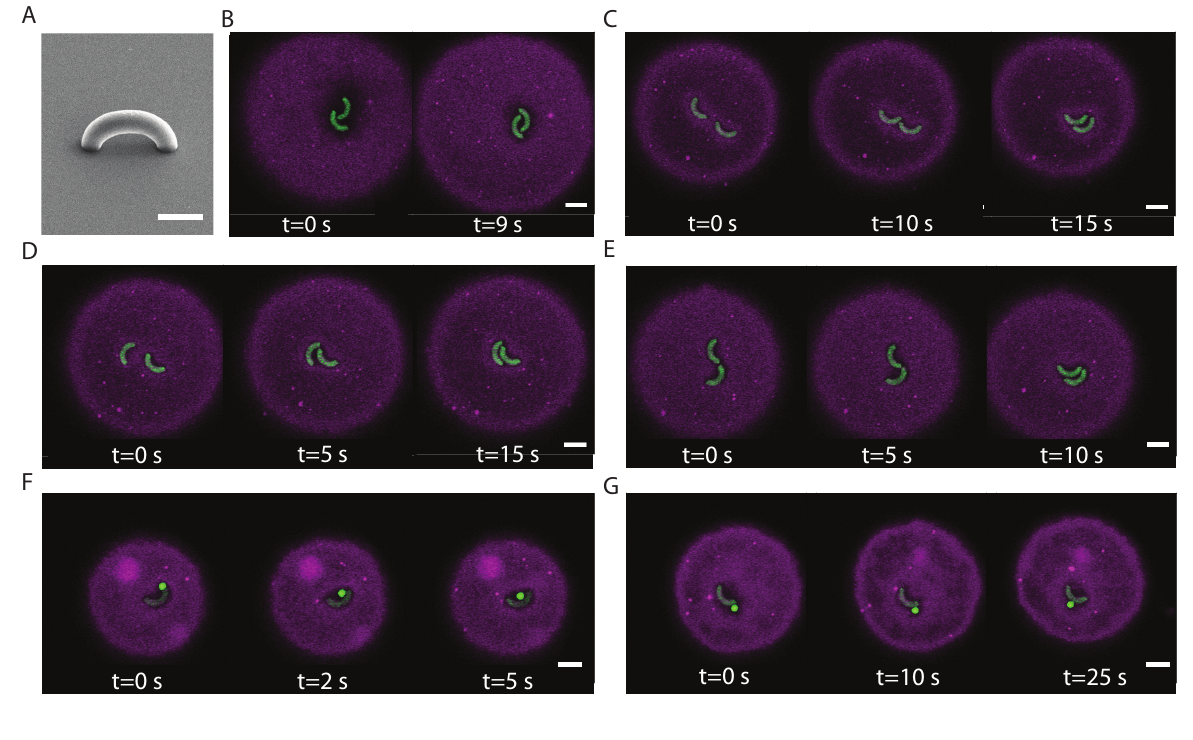}
\caption{\label{fig:crescents}\textbf{Curvature-mediated interaction of bent rod-shaped and spherical colloids:} 
(A) SEM image of a bent rod. 
(B,C,D,E) Time series of confocal images showing the assembly of  two crescents with different initial orientations.  (Vesicle: $h=60$~$\upmu$m and $2R=84$~$\upmu$m). 
(B) when their cavities face each other, they assemble into an interlocked arrangement. 
(C,D, and E) When one of the cavities is closer to the back of the other bent rod than the cavity, they assemble into a stacked configuration. 
(F,G) Time series of confocal images of the curvature-driven assembly  of a bent rod and a sphere (vesicle with $h=36$~$\upmu$m and $2R=48$~$\upmu$m). 
(F) when the sphere is initially located close to the cavity, assembly occurs in a few seconds. 
(G) When the sphere is initially located on the outside of the arc, it migrates towards the center of the outer arc in tens of seconds. See also Supplementary Videos 7 and 8. 
Scale bars 5~$\upmu$m.
}
\end{figure}

Having identified the formation of compact aggregates as a unifying principle for membrane-mediated self-assembly, we now turn to investigating bent rods, a strongly anisotropic shape that combines pronounced curvature with shape complementarity and is reminiscent of BAR-domain proteins.\cite{Peter2004} 
To explore how these geometric properties influence their assembly, we printed curved rods with an opening angle of 180$^\circ$, a diameter of 10~$\upmu$m, and a thickness of 2~$\upmu$m, see Fig. \ref{fig:crescents}A. Unlike other shapes studied so far, the bent rod shape possesses a concave cavity that enables interlocking and stacking useful for compact arrangements.\cite{Riedel2023DesigningParticles} 

Indeed, colloids released from the optical traps with their opposing cavities facing each other never altered their orientation and formed interlocked pairs (Fig. \ref{fig:crescents}B, Supplementary Video 7). In contrast, particles released with one cavity closer to the other's outer arc than to the other's cavity, assembled into stacked configurations (Fig. \ref{fig:crescents}C, D), even when the distance was large, see Fig. \ref{fig:crescents}E and Supplementary Video 7. The shape complementarity thus favors stacked arrangements. 

While we consistently find compact arrangements, the final structure depended on the initial orientation. Similar to the behavior observed for scalene triangles, we find that their initial orientation is important. These observations suggest the presence of different meta-stable arrangements separated by energy barriers that cannot be overcome by thermal fluctuations.

To further probe the role of shape complementarity, we investigated how bent rods interact with 1.25~$\upmu$m diameter spheres that fit into the cavity of the rod. Their shape-complementary fit is reminiscent of the "lock and key" mechanism in proteins. 
Spheres initially positioned near the cavity moved to the center of the arc within a few seconds, as shown in Fig.~\ref{fig:crescents}F. In contrast, spheres initially located on the other side of the bent rods migrated slowly to the center of the outer arc, see Fig.~\ref{fig:crescents}G and Supplementary Video 8. The spheres never passed the tips of the bent rods. These observations indicate that again multiple energy minima exist, separated by an energy barrier imposed by the highly curved tip of the bent rods.

\FloatBarrier
\section*{Conclusion}
Here we experimentally investigate the membrane-curvature mediated interactions between anisotropic colloids for the first time. Using optical tweezers, we placed solid anisotropic colloids obtained from chemical synthesis or 3D printing between a coverslip and a lipid membrane formed by sessile, flattened GUVs. This attachment-free geometry allowed the particles to deform the membrane while remaining free to translate and rotate, enabling direct visualization of their membrane-mediated interactions and assembly pathways.

We studied the behavior of a a large variety of anisotropic colloidal particles, including ellipsoids, dumbbells, cubes, scalene triangles, tetrahedra, and bent rods, and found that they all exhibited membrane-deformation mediated attractions. 
Our observations revealed that particle shape controls both the self-assembly pathway and the final configuration. 
The minimization of the total induced curvature drives this assembly and selects the most compact and spherical arrangement of the particles. As a consequence, the final arrangements favor side-by-side alignment of elongated particles and face-to-face alignment of polyhedra as long as compact arrangement can be maintained. Movement out of the plane is possible if it reduces strong membrane deformations. Regions of high curvature can impose energy barriers that cannot be crossed and lead to long-lived metastable states. Shape complementarity can further be used to guide the assembly into specific, compact configurations.

Our experiments provide a model-system perspective on how anisotropic membrane-deforming objects organize through the deformations they generate. The observed tendency of particles to align side-by-side and face-to-face might be important for structures such as actin filaments, microtubuli or protein assemblies that impose anisotropic curvature on lipid membranes. More broadly, our results establish particle shape as a key parameter for membrane-mediated self-assembly, controlling the interactions, assembly pathways, orientational rearrangements, and kinetic trapping. 

Future work could extend this approach to more complex lipid membranes, for example multicomponent membranes that feature two or more phase-separated domains that are characterized by distinct order and disorder phases and thus possess different stiffness~\cite{Veatch2003,Needham1988}.  
This would introduce an additional layer of complexity akin to biological membranes.

\section*{Material and Methods}
\subsection*{Materials}

Sucrose, D-glucose, Hydrochloric acid (\ch{HCl}) 37\%, Acetic acid, 3-(trimethoxysilyl)propyl methacrylate (TPM), N,N,N',N'-tetramethylethylenediamine (TEMED), ammonium persulfate (APS) and Chloroform (\ch{CHCl3}) 99.0-99.4\%
were provided from Sigma-Aldrich. 
DOPC ( $\Delta$ 9-cis 1,2-dioleoyl-sn-glycero-3-phosphocholine), 
DOPE-rhodamine  (1,2-dioleoyl-sn-glycero-3-phosphoethanolamine-N-(lissaminerhodamine B sulfonyl)), 
DOPE-PEG2000 (1,2-dioleoyl-sn-glycero-3-phosphoethanolamine-N-[(polyethylene glycol)-2000]), and TopFluor Cholesterol were purchased from Avanti Polar Lipids. 
Sodium azide  (\ch{NaN3}) 99\% extra pure, and Potassium chloride (\ch{KCl}) 99+\%
were purchesed from Acros Organics. 
Phosphate buffered saline (PBS) tablets from Merck Millipore. Ethanol (\ch{C2H5OH}) from Honeywell, and water (\ch{H2O}) was filtered with a MilliQ MilliPore apparatus (resistivity 18.2 M$\Omega\cdot$ cm). 
All chemicals were used as received.

\subsection*{Coverslip Functionalization}
To functionalize the coverslips, we followed a previously described protocol \cite{VanDerWel2016, Azadbakht2023a,melio_pivoting_2026}. Briefly, we started by sonicating the coverslips in 1M KOH solution, then rinsed them by ethanol and three times by water. Next, they were submerged in ethanol with 1\%v/v acetic acid and 0.5\%v/v TPM for 15 minutes. After functionalization, they were rinsed three times with ethanol and baked in a preheated oven for 1 h at 80$^{\circ}$C. The polymerization of acrylamide onto the TPM-coated glasses was performed for two hours using a 2\% w/w acrylamide solution that had been degassed in a vacuum for 30 minutes to remove oxygen. This solution was combined with 0.035\% v/w TEMED and 0.070\% w/w APS to initiate the reaction. The resulting functionalized coverslips were subsequently incubated in a refrigerated environment at 4$^{\circ}$C while immersed in the polymerization solution. Immediately before use, the coverslips were thoroughly rinsed with water and blow-dried with nitrogen.

\subsection*{Printed colloids}

A 3D microprinter (Photonic Professional GT, Nanoscribe GmbH) was utilized to print colloids using two-photon polymerization (2PP) at the micrometer scale. The printer was equipped with a 63$\times$ oil-immersion objective lens (Zeiss, NA = 1.4) for the highest optical resolution. Arrays of colloids were designed using Autodesk Inventor software and then processed using Describe. The printing was performed in dip mode using a commercial photoresist called IP-Dip (Nanoscribe GmbH).~\cite{Doherty2020}.

Once the printing process was completed, the substrate with the printed structure was immersed in propylene glycol methylether acrylate (PGMA) for 30~min and then immersed in isopropanol for 1~min for development. The substrate was kept in a Petri dish overnight to evaporate the remaining organic solution. To detach the particles from the substrate, the substrate was immersed in a 300~mM glucose solution and sonicated for 10 minutes. 
After detachement of the particles, the substrate was removed and the particle solution was harvested and centrifuged to concentrate the particles by removing the supernatant. 


\subsection*{Synthetic Colloidal particles}

Dumbbell-shaped and cubic particles were made of iron oxide, following ref.\cite{Wang2013,Sugimoto1992,Shelke2023ExploitingPurity}. 
After coating the particle with silica through a St\"ober process following ref.~\cite{Wang2013,Rossi2011CubicColloids}, the templating hematite core was removed by hydrochloric acid (HCl) to prevent laser-induced heating during optical tweezers manipulation. Subsequently, the colloids were washed with water three times.

To facilitate imaging with a confocal microscope and quantification of their position and orientation, the hollow silica was further coated with a fluorescently labelled lipid bilayer containing TopFluor Cholesterol (23-(dipyrrometheneboron difluoride)-24-norcholesterol, following the protocol described in \citenum{melio_pivoting_2026}. 
The first step was to dry the lipid mixtures composed of 97.5 wt\% DOPC, 2 wt\% DOPE-PEG2000, and 0.5 wt\% TopFluor Cholesterol in vaccum. Then HEPES buffer (10~mM~HEPES, 50~mM~NaCl, at pH=7.00) was added, and the lipid mixture was extruded through two polycarbonate filters with pore size of 30 nm for 21 times, to produce small unilamellar vesicles (SUVs). To coat the hollow silica particles, they were mixed with SUVs by placing them on a rotator at 5~RPM for 2~hour. Finally, the particles were washed three times with HEPES buffer.

Polystyrene spheres were synthesized with BODIPY fluorescent dye with a diameter of 1.25 $\upmu$m and polydispersity of 5.75\% as described in ref. \cite{Appel2013,VanDerWel2016}

\subsection*{Giant unilamellar vesicle (GUV) fabrication}
GUVs were prepared by a standard electroformation method, yielding vesicles 10 to 100~$\upmu$m in diameter~\cite{Dietrich1997}. 
The GUVs were composed of 1,2-dioleoyl-sn-glycero-3-phosphocholine (DOPC) containing a 2\% w/w polyethylene glycol-2000 conjugated to 1,2-dioleoyl-sn-glycero-3-phosphoethanolamine (DOPE-PEG200) to suppress nonspecific attachment, and 0.5 wt\% DOPE-rhodamine was dried on two ITO-coated glass slides in a vacuum chamber for more than 2 hours. ITO-glasses were immersed in a 300mM sucrose solution, and a AC voltage of 2.1 V(rms) was applied at 10~Hz for 2~hours. Then the frequency was decreased to 6~Hz, resulting in larger vesicles.

\subsection*{Sample Preparation}
GUVs were electroswollen with 300~mM sucrose solution which was encapsulated on the inside of the GUVs. As outer solution, a 308~mM mixture consisting of 80\% vol of an aqueous 310~mM glucose solution and 20\% vol of 150~mM NaCl solution was used. The density and osmolarity differences between the inner and outer solution of $\Delta \rho$ = 22.6~kg/m$^3$ and 8~mM, respectively, caused the GUVs to deflate and sediment to the polyacrylamide-coated coverglasses. Experiments were conducted at ambient room temperature. The polyacrylamide prevented nonspecific adhesion of the GUV membrane and particles to the glass (see Methods).

\subsection*{Microscopy and Optical Trapping}

An A1R Nikon confocal scanner was used to capture images on a Ti-E microscope, with a 60$\times$ water immersion objective (N.A.~=~1.2). The confocal scanner was operated in resonance mode, producing 30~frames/s of 512~$\times$~512 frames with 102~px/nm.

Vesicles were doped with 0.5~wt\% rhodamine and excited with a 561 nm laser, with emission light collected in the range of 570-620~nm. Spheres were stained by BODIPY, CSLBs contained 0.5 wt\% TopFluor, and printed colloids were auto-fluorescent as is. All colloids were excited with a 488nm laser, with emission light collected at 500-500 nm. Two separate photodetectors were used to simultaneously detect the colloidal particles and the GUVs. 

A Meadowlark Spatial Light Modulator (SLM) (HS1920) was used to modulate the phase of the laser wavefront. A 1064 nm laser beam from LaserQUANTUM was expanded to fill the aperture of the SLM and was imaged onto the back focal plane of the microscope objective using two plano-convex lenses. The non-diffracted light was filtered out in the front focal point of the first lens. The light was then directed into the light path of the microscope via a dichroic mirror to allow for simultaneous imaging and trapping.

Red Tweezers software\cite{Bowman2014a} was used to program the SLM, running on a Geforce RTX4000 Graphics processing unit. This setup generated 100 holograms per second while the SLM refreshing speed was 120~Hz.

Scanning Electron microscopy (SEM) images were captured using a Thermo-Fisher ApreoSEM. The imaging was performed on two different samples: (1) an array of particles immediately after printing and sputter-coating of 5~$n$m platinum / palladium 80:20, and 
(2) a particle dispersion dried on a SEM stub. 

\subsection*{Energy Minimization}

The membrane shape is described in Monge parametrization by the height $h(x,y)$ above a reference $x$-$y$ plane, i.e., the midplane of the flat membrane on the substrate away from the particle positions. In this parametrization, the mean
curvature of the membrane shape can be expressed as
\begin{equation}
H = \frac{(1 + h_{x}^2) h_{yy} + (1 + h_{y}^2) h_{xx} - 2 h_{x} h_{y} h_{xy}}{2 (1 + h_{x}^2 + h_{y}^2 )^{3/2}}
\end{equation}
with subscripts $x$ and $y$  to determine the
membrane bending energy described by the Helfrich Hamiltonian
\begin{equation}
E_{\rm be} = \int 2 \kappa H^2 {\rm d}A
\end{equation}
with bending rigidity $\kappa$ and membrane area element 
\begin{equation}
{\rm d}A = \sqrt{1 + h_{x}^2 + h_{y}^2}\;{\rm d}x\,{\rm d}y
\end{equation}
In the numerical energy-minimization approach, the reference plane is discretized  into a square lattice with lattice constant 200 nm, which is much smaller than the particle dimensions. Partial derivatives in the bending energy are approximated as standard multivariate finite differences. To constrain the confinement volume $v$ per particle, a pressure $p$ is used as Lagrange multiplier in the energy $E = E_{\rm be} + p V$ with $V = n_p v =  \int h\; {\rm d}x\,{\rm d}y$, where $n_p$ is the number of particles.  The total confinement volume $V$ and and bending energy $E_{\rm be}$ then are determined in energy minimizations and interpolated to values of $p$ at which the desired value of $v$ is obtained. All minimizations are performed with the function FindMinimum of Mathematica 13.3 \cite{Mathematica}. In these minimizations, the position and spherical shape of the particles are taken into account by constraints on the height $h(x,y)$, i.e.~by lower bounds for $h$ at lattice sites located under the particles to prevent an overlap of the membrane with the particles.

\section*{Acknowledgment}

We thank Yogesh Shelke and Julio Melio for preparing PAA coverslips and synthesizing dumbbells colloidal particles as well as Rachel Doherty and Solenn Riedel for help with printing the particles.

\section*{Competing interests}
The authors declare no competing interests. 

\bibliography{references_all}
\clearpage


\section*{SUPPORTING INFORMATION}

\renewcommand{\thefigure}{S\arabic{figure}}
\renewcommand{\thetable}{S\arabic{table}}
\setcounter{figure}{0}

\section*{Supporting Figures}
\begin{figure}[ht]
\centering
\includegraphics[width=0.6\textwidth]{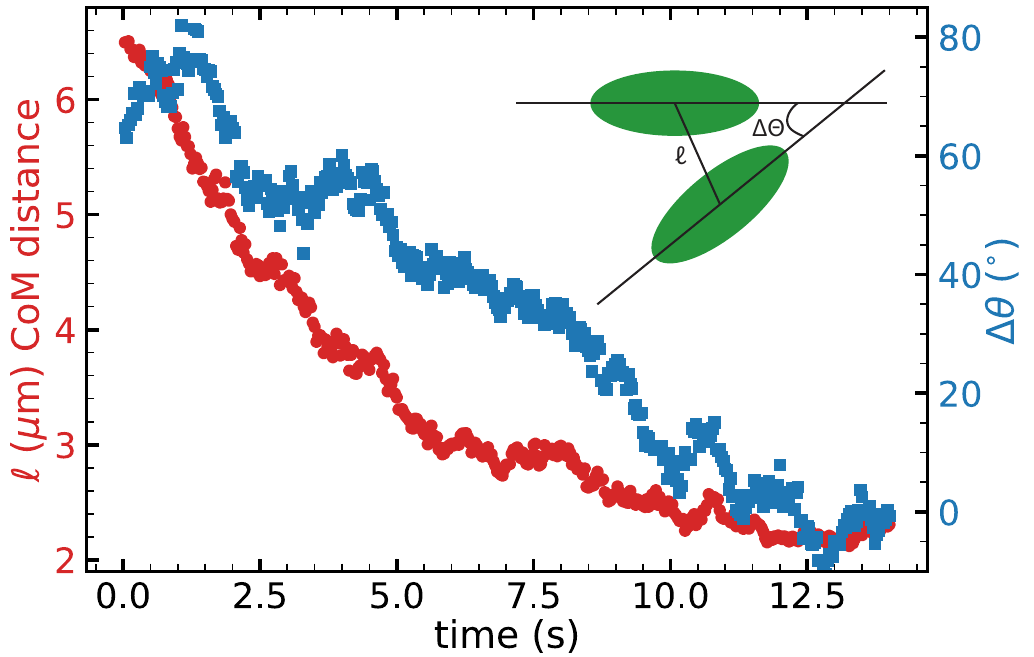}
\caption{\label{fig:SI_ellipse} \textbf{Quantification of two approaching ellipses.}
(red circles, left axis) Distance of the center-of-mass (CoM) ($\ell$) and (blue squares, right axis) angle between the long axes between two membrane-deforming ellipsoids as a function of time. \\
Particle tracking was performed by a series of image processing steps. First, a Gaussian blur with a kernel size of 1 pixel was applied to improve the image quality, followed by ann intensity threshold to identify distinct particles in the processed image. The OpenCV package in Python was used to detect contours outlining the boundaries of the particles. Subsequently, the contour of each particle was approximated using an ellipse to capture the shape features such as the center-of-mass and orientation.}
\end{figure}

\begin{figure}[ht]
\centering
\includegraphics[width=1\textwidth]{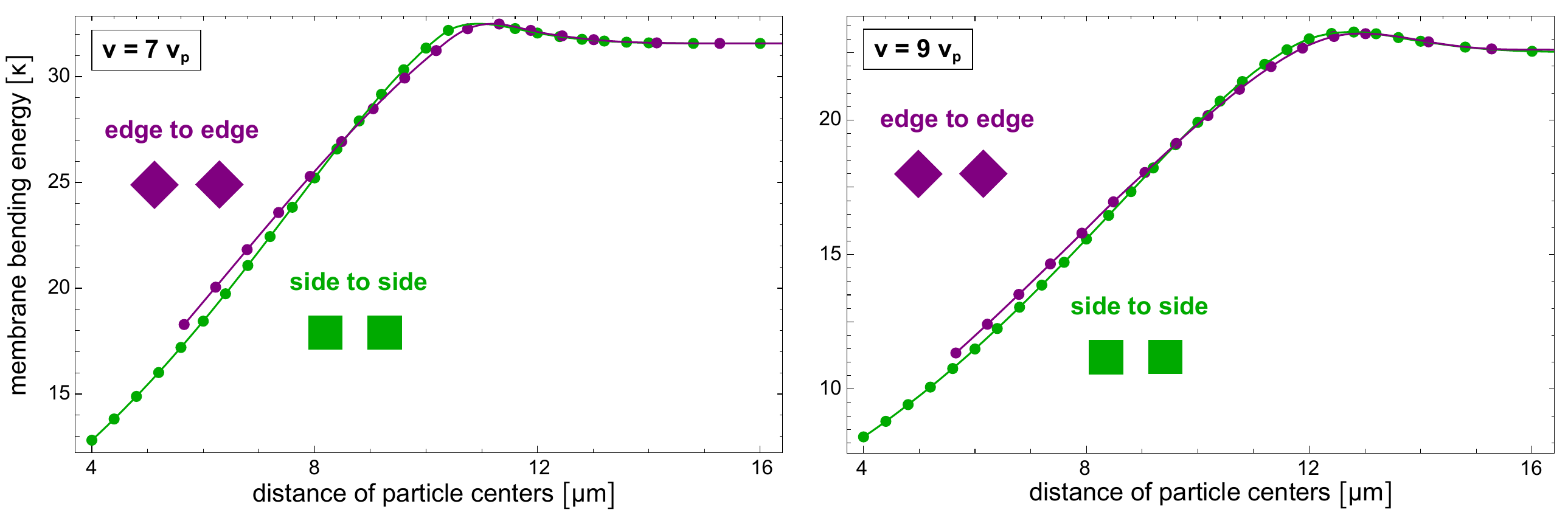}
\caption{\label{fig:SI_cubes}\textbf{Interaction energies of membrane-confined cubes:} 
 Membrane bending energy versus distance of particle centers for side-to-side and edge-to-edge orientation of two cubic particles with a length of 4 $\upmu$m  at confinement volume values $v = 7 v_p$ and $9 v_p$ where $v_p$ is the particle volume. The minimum distances of the cubic particles in side-to-side and edge-to-edge orientation are 4 $\upmu$m and $4\sqrt{2}$ $\upmu$m $\simeq$ 5.66 $\upmu$m.  At intermediate distances between about 8.5 and 11 $\upmu$m for $v = 7 v_p$ and between about 10 and 13 $\upmu$m for $v = 9 v_p$, the membrane bending energy is lower for an edge-to-edge orientation of the cubes, by up to about 0.5 $\kappa$ for $v = 7 v_p$ and 0.25 $\kappa$ for $v = 9 v_p$. At smaller distances, the membrane bending energy is lower in side-to-side orientation.
 }
\end{figure}

\begin{figure}[ht]
\centering
\includegraphics[width=1\textwidth]{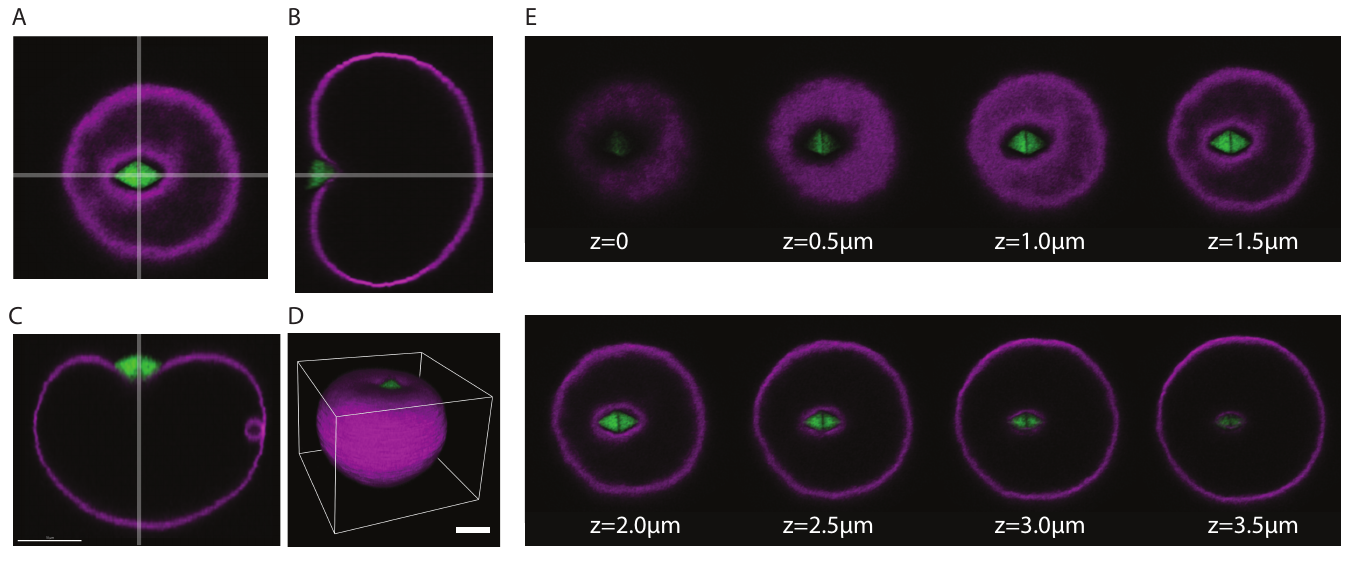}
\caption{\label{fig:SI_tetra}\textbf{Curvature-mediated interaction of two tetrahedral colloids.} (A-C) Slices and (D) 3D reconstructed image of the final arrangement of two membrane-confined tetrahedral particles. 
(A) a confocal image of the focal plane close to the coverslip;
(B) a zy-plane cross section made along the vertical line in (A).
(C) a xz-plane cross section taken along the horizontal line in (A).
(E) Z-stacks of confocal microscopy images take at different heights as indicated.
}
\end{figure}

\end{document}